\documentstyle[aps,prl,floats,epsf,amssymb]{revtex}

\begin{document}
\draft
\wideabs{
\title{Depinning with Dynamic Stress Overshoots: Mean Field Theory}
\date {December 13, 2000}
\author{J.M. Schwarz and Daniel S. Fisher}
\address{Lyman Laboratory of Physics, Harvard University, Cambridge, 
Massachusetts 02138}
\maketitle

\begin{abstract}
An infinite-range model of an elastic manifold
pulled through a random potential by an applied force $F$ is analyzed 
focusing on inertial effects.  When the inertial parameter, $M$, is small, there is a continuous depinning transition from a small-$F$ static phase to a large-$F$ moving phase.  When $M$ is increased to $M_c$, a novel tricritical point occurs.  For $M\!>\!M_c$, the depinning transition becomes discontinuous with hysteresis.  Yet, the distribution of discrete ``avalanche''-like events as the force is increased in the static phase for $M\!>\!M_c$ has an unusual mixture of first-order-like and critical features.
\end{abstract}


}

A wide variety of driven systems are characterized by some type of elastic,
extended object that is pulled through a quenched 
random medium by a uniform 
applied force.  If the dynamics is dissipative-- e.g. sliding charge
density waves [1], vortex lattices in superconductors [2] and domain walls
in ferromagnets [3] --such systems exhibit
a critical depinning transition in the absence of thermal fluctuations.  
A small applied
force, $F$, is not enough
 to overcome
the random pinning forces and the system remains trapped in 
one of many possible metastable configurations.  
However, as $F$ is slowly increased, some sections
become unstable and move, only to be stopped by the  
elastic forces from
more strongly pinned neighboring regions. As $F$ is increased even further,
there will be a sequence of these discrete, localized ``avalanche'' events with
a distribution of sizes, $s$ [4].  As a critical force $F_c$ is approached
from below,
arbitrarily large avalanches can occur.   Above $F_{c}$, the 
driving force is able to overcome the pinning, giving rise to a non-zero
average velocity, $\bar{v}$.  The transition between the two phases
is continuous with $\bar{v}$ playing the role of an order parameter that 
vanishes as $\bar{v}\sim(F-F_{c})^{\beta}$ as
$F_{c}$ is approached from above.  This class of dissipative systems
has been analyzed by renormalization group (RG) methods [1]
with the results supported by numerical [5] and   
experimental [6] evidence, although quantitative experimental tests have been very limited.

In some systems, in particular crack fronts in brittle materials [7] and motion
of contact lines of droplets on dirty or rough solid surfaces [8], the dynamics are
not dissipative and inertial effects can be important.  In this Letter, 
we explore the effects of inertial stress transfer in which the motion of one segment creates a
transient stress (in addition to the static elastic stress)
on other segments.  In the presence of pinning, the
motion will
be mostly forward; this makes the positive (forward pulling) parts of the
transient stress transfer the most important; we call these
\emph{stress overshoots}.  Although we will focus only on 
such dynamic stress transfer effects here, the more direct effects of
inertia will have similar consequences: if a segment moves forward
in an underdamped manner so that it overshoots -- passing a new local minimum 
before relaxing back into it -- the stress it transfers to the other regions
will reflect this overshoot.  

Our focus will be on the nature of the
depinning transition in the presence of stress overshoots; a key 
question is whether or not the transition from the pinned to the 
moving phase remains continuous or becomes ``first-order''-- 
characterized by a finite jump in $\bar{v}$ as a function of the
applied force and perhaps by hysteresis.  We will also consider the 
distribution of avalanche sizes as the depinning transition is
approached from below.  

Near the depinning transition, the motion will be very jerky with a lot of
starting and stopping.  Thus, the essential physics can be 
captured by models in which space, time and the manifold's displacement in the
direction of motion, $h({\bf x},t)$, are all discrete.  The elastic stress on
a segment ${\bf x}$ takes the general form
\begin{equation}
\sigma({\bf x},t)=\sum_{{\bf y}}\sum_{\tau\ge\,0}J_{{\bf x}{\bf y}}(\tau)h({\bf y},t-\tau)-\hat{J}\,h({\bf x},t)\,,
\end{equation}
with the static stress transfer given by $J^{s}_{{\bf x}{\bf y}}=\sum_{\tau}J_{{\bf x}{\bf y}}(\tau)$
and $\hat{J}=\sum_{{\bf y}}J_{{\bf x}{\bf y}}^{s}$.  The simplest form 
of the stress overshoots is when they only last for
one time step, i.e. $J_{{\bf x}{\bf y}}(\tau\ge\,2)=0$, and $J_{{\bf x}{\bf y}}$ is non-zero only for
nearest neighbors.  In this 
case we write, $J_{{\bf x}{\bf y}}(0)=
\frac{(1+M)}{Z}$ and $J_{{\bf x}{\bf y}}(1)=
-\frac{M}{Z}$ for ${\bf x}\ne {\bf y}$ where $Z$ is the number of nearest neighbors and $M$
is the magnitude of the stress overshoot.  With this dynamic stress
transfer, the motion
of any neighboring segment, ${\bf y}$, causes an extra stress on the ${\bf x}{\scriptstyle  th}$ segment 
for one time step only.  For
$M<0$, a key property of the elastic stress transfer is that it is
monotonic, i.e. increasing in time when the $h$'s increase.  This 
monotonicity property can be used to prove the uniqueness of the 
average velocity and thereby a unique critical force [9].  It is also an 
essential feature for the validity of the RG analysis [1].  For $M>0$, 
the stress transfer is {\it non}-monotonic in
time so that the average velocity is not necessarily a unique function 
of $F$.

We take the displacements $h({\bf x},t)$ to be pinned at
a discrete set of possible values with associated pinning (yield) strengths 
$f_{Y}[{\bf x},h({\bf x},t)]$.  The dynamics
is simple: if the total force on segment
${\bf x}$ exceeds $f_{Y}$ at that point, the segment jumps forward, i.e.
if $\sigma({\bf x},t)+F>f_{Y}[{\bf x},h({\bf x},t)]$ then $h({\bf x},t+1)=h({\bf x},t)+\Delta[{\bf x},h({\bf x},t)]$ 
with $\Delta$ the distance to the next pinning position; otherwise 
$h({\bf x},t+1)=h({\bf x},t)$.   As long as the jumps
$\Delta[{\bf x},h({\bf x},t)]$ are random variables independently drawn from 
a smooth distribution $D(\Delta)d\Delta$, the randomness in the pinning 
strengths is not essential and, for simplicity, we make them all equal. 

We now take the first step in analyzing such systems by studying a mean
field-- more precisely an infinite-range --model.  For the case without
stress overshoots, such a model was the needed starting point for the
RG analysis of finite-dimensional systems.  We thus consider a 
model of $N$ segments each coupled to all the others.  The stress on 
$h(x,t)$ is then
simply  
\begin{equation}
\sigma(x,t)=\bar{h}(t)-h(x,t)+M(\bar{h}(t)-\bar{h}(t-1))\,,
\end{equation}
where $\bar{h}(t)\equiv\frac{1}{N}\sum_{y}h(y,t)$ is the average over the
other segments.  

In this infinite range model, all sites are statistically equivalent and 
coupled only via the mean field, $\bar{h}(t)$.  We can therefore fully characterize
the configuration by 
the distribution of the excess force on a 
segment $f_{t}(x)\equiv\sigma(x,t)+F-f_{Y}$.
Dropping the $x$ index, we denote this distribution, $\rho_{t}(f_{t})df_{t}$.
Since each segment is either pinned, $f_{t}\leq 0$, 
or will jump forward by some
random amount, $\Delta$, the equation of motion for the
excess force distribution is 
\begin{eqnarray}
\rho_{t+1}(f_{t+1})\!=\!\int^{\infty}_{-\infty}\!\!\!d\phi_t\delta(\phi_t\!+\!\bar{v}_
{t+1}\!+\!M(\bar{v}_{t+1}\!-\!\bar{v}_t)     
\!-\!f_{t+1}) \nonumber \\
\times\{\rho_t(\phi_t)\Theta(-\phi_t)\!+\!\int_{0}^{\infty}\!\!\!\!D(\Delta)\rho_{t}(\phi_t+\Delta)\Theta(\phi_{t}+\Delta)d\Delta\},
\end{eqnarray}
where $\phi_t$ is the new excess force on a segment \emph{before} the others have
moved,
$\bar{v}_{t}\equiv\bar{h}(t)-\bar{h}(t-1)$ 
and $\Theta$ is the unit step function.  
The self-consistency relation for $\bar{v}_{t}$ follows from the
``spatial'' averaging of the excess force:
\begin{equation}
<f_{t}>=\int_{-\infty}^{\infty}f_{t}\rho_{t}(f_{t})df_{t}=F -f_Y
+ M\bar{v}_{t}\,.
\label{4} \end{equation} 
For static solutions, $\bar{v}=0$, the stress overshoot plays no role and one can show that there are many solutions 
as long as the applied force is less than the static critical force
$F_{c0}=f_Y-\frac{1}{2}\frac{\overline{\Delta^{2}}}{\bar{\Delta}}$, where
the bars here denote averages over $D(\Delta)$.  We will return later to the response to an increase in the force in this regime. For now, we turn to the moving phase.

We first consider the steady state limit for which 
$\bar{v}_{t}=\bar{v}>0$ is independent of time.  Equation (\ref{4}) implies
that in this case the stress overshoot has the same effect as an
additional applied force, $F\rightarrow\,F+M\bar{v}$, and the steady state
$\rho(f)$ depends only on $\bar{v}$ (and $D$).  
It is easiest to work with the particular jump distribution
$D(\Delta)=e^{-\Delta}$ which corresponds to random pinning positions
with density $1/\bar{\Delta}$ chosen to be unity.  However, 
the qualitative
properties of the $\bar{v}(F)$ curve are applicable to any well-behaved,
smooth pinning distribution with $D(0)$ finite and non-zero.  For $M<1$ and 
$F>F_{c0}$,
$\bar{v}$ rises
continuously as
\begin{equation}
\bar{v}\sim\frac{F-F_{c0}}{1-M}
\end{equation}
for $\bar{v}<<\bar{\Delta}=1$.   The fact that $\bar{v}$ is linearly proportional
to $F-F_c$ in mean field models with jumps has been
derived previously for dissipation-dominated systems[1]; 
it is due to the fact that above the depinning 
transition the average displacement of a segment responds linearly in a 
non-singular way to an increase in the \emph{total} force
on it.   Note that as the applied force is increased further, the $\bar{v}$ of
the model is constrained in an unphysical way by the average distance
between pinning sites.  As a result, $\bar{v}$ 
will saturate in a manner that depends on
the particulars of the pinning distribution.  For the exponentially decaying
jump distribution, the full expression for $\bar{v}$ is
\begin{equation}
\bar{v}=\frac{-\epsilon_{0}+\mu+\sqrt{(\epsilon_{0}+\mu)^{2}+2\epsilon_{0}}}
{1+2\mu}
\end{equation}
with $\epsilon_{o}\equiv\,F-F_{c0}$ and $\mu\equiv\,M-M_{c}$. 

As $M$  increases, the amount
of applied force required to reach a specific $\bar{v}$ decreases.  When 
$M$ reaches a critical value of $M_{c}=1$, the width of the continuous
depinning transition shrinks to zero.  When both the parameters $M$ and
$F$ are critical there is a ``tricritical'' point. Near this 
tricritical point the  scaling behavior is 
$\bar{v}\sim|\mu|\sim\sqrt{\epsilon_{0}}$, indicating a new
universality class of depinning transitions.  
 
For $M>M_{c}$, there is a {\it jump} in $\bar{v}$ from 
$\bar{v}_{min}\sim \mu$ to zero when $F$ is decreased
through a new {\it lower critical force} $F_{c}^{\downarrow}=F_{c0}-(1+\mu-
\sqrt{1+2\mu})$.   The two stable solutions in 
this regime consist of $\bar{v}=0$ and a 
branch with $\frac{d\bar{v}}{dF}<0$.  There is
also a third solution, but it is unstable.  
Between $F^{\downarrow}_{c}$ and $F_{c0}$ there is hysteresis as might be
expected at a first-order like phase transition.   A greater 
decrease in the force is required to stop the 
motion in the presence of overshoots
than in their absence, but once the overall motion does stop the
overshoots will have less effect as $F$ is increased back up again.  

We now investigate the avalanche dynamics below the depinning
threshold.
An avalanche is the forward motion of a finite number of 
segments in response to the
initial motion of one segment.  The applied force is increased only
in order to trigger the motion of the initial segment and is then held
fixed until the avalanche stops.  The segments that move forward in
a single finite avalanche are those whose $f$ is within an infinitesimal--
$\mathcal{O}$$(\frac{1}{N})$--
slice near $f=0$ of the distribution $\rho(f)$; for 
continuous $\rho(f)$ all that matters
is thus $\rho(0)$.  The total number of segments that hop forward at one
time step, $n_{t}$, will have poissonian statistics whose mean is 
determined by the increase in the total force on a segment from the
previous time step.  
The mean number that hop at a given time $t$ is
given by 
\begin{equation}
<n_{t}>=\rho(0)\bar{\Delta}[(1+M)n_{t-1}-M\,n_{t-2}]\,.
\end{equation}
If this mean becomes negative, the avalanche will stop.  We 
define the parameter, $\Gamma\equiv\,\rho(0)\bar{\Delta}$, which is 
essentially the mean local response to an increase in the total force.

We are primarily interested in the large avalanches, for which $n_{t}$
will typically be large.  Given large $n_{t-2}$ and $n_{t-1}$, the 
randomness will cause approximately
gaussian fluctuations in $n_{t}$  of magnitude of order $\sqrt{n_{t}}$.
With no stress overshoots, $M=0$, we can approximate the dynamics in
continuous time in the Ito representation as
\begin{equation}
\frac{dn(t)}{dt}\simeq(\Gamma-1)n(t)+\sqrt{n(t)}\eta(t)
\end{equation}
with $\eta$ white noise.  The constraint that $n(t)$ be positive makes 
finding the distribution of the avalanche size, $s\equiv\int dt\,n_{t}$, 
a first passage problem.  
A critical point occurs at $\Gamma=1$ 
($\rho_{c}(0)=\frac{1}{\bar{\Delta}}$), at which there is a 
power law distribution of avalanche sizes, ${\rm Prob}(ds)\sim \frac{1}
{s^{\frac{3}{2}}}ds$ for large $s$ [10].  Beyond this point, infinite
avalanches occur as the motion of one segment spawns more than 
one descendant at the next time step.    
For smaller $\Gamma$, a cutoff
in the avalanche size distribution given by $\tilde{s}\sim\frac{1}
{(1-\Gamma)^{2}}$
( which is not equal to the average as the distribution is very broad), and
${\rm Prob}(ds)\sim\frac{1}{s^{\frac{3}{2}}}e^{-\frac{(1-\Gamma)^{2}s}{4}}ds$ 
for large $s$ as shown in Fig. 1.  

In the more general case with stress overshoots, the stochastic equation
becomes second-order in time since the number of segments that hop
forward is dependent on two previous time steps and we have, approximately,
\begin{eqnarray}
\Gamma M\frac{d^{2}n(t)}{dt^{2}}\simeq(\Gamma-1)n(t)+ 
(\Gamma\,M-1)\frac{dn(t)}{dt}   \nonumber \\
+\sqrt{n(t)}\eta(t)\,.
\end{eqnarray}
When we introduce a {\it small} amount 
of stress overshoot, the avalanche dynamics
is strongly overdamped.  The critical point is still at $\Gamma=1$ but
the distribution of avalanche sizes is 
modified by the coefficient $(M-1)$. 

 When
the stress overshoot magnitude is increased to the critical value, 
$M_{c}=1$, the avalanche dynamics becomes undamped resulting in
a new exponent for the distribution of avalanche sizes.  At this tricritical
point, we have $\frac{d^{2}n(t)}{dt^{2}}=\sqrt{n(t)}\eta(t)$ which 
maps the distribution of avalanche sizes 
into a first passage problem of a particle with random acceleration.
The distribution of avalanche sizes broadens to ${\rm Prob}(ds)\sim \frac{1}
{s^{\frac{5}{4}}}ds$ for large $s$ [11].  
The typical avalanche of size $s$ now lasts
for a time $\tau\sim\,s^{\frac{1}{4}}$ in contrast to $\tau\sim\,s
^{\frac{1}{2}}$ in the absence of stress overshoots.  The scaling form
of the distribution within the vicinity of this tricritical point is
\begin{equation}
{\rm Prob}(ds)\sim\frac{1}{s^{5/4}}F(s(M-M_c),
s^{1/2}(1-\Gamma))ds.
\end{equation}
    
For $M>M_{c}$, the deterministic
part of the avalanche equation of motion is that of a time-reversed 
harmonic oscillator.  If $\rho(0)$ is small-- i.e., below the critically 
damped line --$n(t)$ grows exponentially with time before crashing down
to zero  as a result of the oscillation.  Above
some critical $\Gamma$, $\Gamma_{c}(M)
=\frac{4M}
{(1+M)^{2}}$ (from Eq. (7)), large 
avalanches grow exponentially and become
infinite.  On the critical line, some avalanches become infinite while
others remain finite.  Because of the exponential growth of $n(t)$ both
above and below the critical line, the randomness will not have much of
an effect once an avalanche becomes large and a deterministic analysis
where all the randomness is encoded into random initial $n$ and $\frac
{dn}{dt}$ becomes valid.
Just below the critical line, there is a clear delineation in the 
distribution of avalanche sizes between those avalanches that would
remain finite and those that would become infinite at the critical 
line.  Indeed, just below $\Gamma_{c}(M)$, a finite fraction of
the avalanches will have size of order $\hat{s}\sim\,exp(A(M)/
\sqrt{(\Gamma_{c}(M)-\Gamma)})$ to within a multiplicative factor
of order unity; these last roughly half the oscillation period.  Note that for
$M\gtrsim 1$, $A\sim\mu\equiv M-1$ and the scaling of $1-\Gamma$ versus $\mu$
is like that of $F-F_{c0}$ versus $\mu$ in the moving phase.  Most of the
other avalanches will be small, but with a long tail in log $s$ scaling as
\begin{equation}
{\rm Prob}(ds)\sim\frac{1}{s(\ln s)^{2}}ds=\frac{d(\ln s)}{(\ln s)^2} 
\end{equation}
for $1<<\ln s<<\ln\hat{s}$.  This is an unusual distribution with
$\ln s$ being the natural variable.  This broad distribution  for the finite avalanches persists
at the depinning transition, but now
there is also a non-zero probability that an avalanche will be infinite 
as indicated in Fig. 1.  These results suggest a new 
hybrid transition for $M>M_c$ 
with a diverging scale,
$\hat{s}$, but also discontinuities, particularly in $\bar{v}(F)$.  
\begin{figure}[t]
\vspace*{-1.3cm}
\begin{center}
\epsfxsize=6.75cm
\epsfysize=6.75cm
\epsfbox{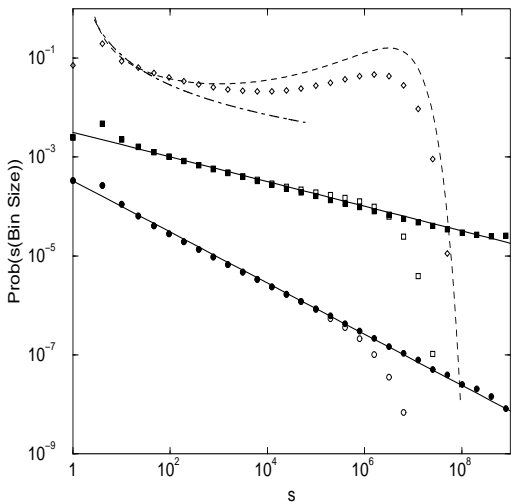}
\caption{Log-log plot of the probability of an avalanche occurring of size 
within the interval [$s/\sqrt{2},\sqrt{2}s$).  The circles
are numerical data in the absence of stress overshoots,
with solid circles at the critical $\Gamma=1$ and open circles just
below, $\Gamma=0.999$.  The squares are at the critical
$M=M_c=1$.  The closed squares are at the tricritical point $\Gamma=1$ 
with the open squares just below, $\Gamma=0.999$.  The solid lines have slope
$\frac{1}{2}$ and $\frac{1}{4}$, representing theoretical 
predictions for the critical $M<1$ and tricritical $M=1$ points.  The open
diamonds are for $M=2$ and $\Gamma=0.88$, just
below the ``first-order'' line; these should roughly match the dashed
curve which is from the deterministic approximation with an exponential
distribution of initial velocities.  The dot-dashed curve represents 
Eq. (11). 
}
\end{center}
\vspace*{-0.75cm}
\end{figure}

So far, we have not established the connection between the driving
force, $F$, and the density of about-to-jump segments $\rho(0)$. 
This is needed to connect together the analysis of the pinned and the
moving phases.  As long as the avalanches remain finite and the
driving force is adiabatically increased,  the total force distribution
$\rho(f)df$ will be
independent of $M$; the avalanche distribution will only affect {\it when}
segments jump and hence the small scale ($f\sim \frac{1}{N}$) behavior
of the stationary configurations between avalanches.  
For generic
initial conditions --- continuous $\rho(f)$ 
which must be consistent with randomly 
positioned pinning values of $h(x,t)$ --- $\Gamma$ will
increase with $F$ to reach unity only right at $F_{c0}$, although
the way in which it approaches this value is non-universal and history
dependent.  Thus for $M<1$, $\Gamma_{c}=1$ and 
the infinite avalanches will only occur when
$F>F_{c0}$, the same point above which $\bar{v}>0$.  

For
$M>1$, however, $\Gamma$ will reach the value $\Gamma_{c}(M)$ at a value
of the driving force $F^{\uparrow}_{c}$, 
that depends on the initial conditions as well
as on $M$.  From an analysis of Eq(3), it can be shown that $F^{\uparrow}_{c}$
is between $F_{c}^{\downarrow}(M)$, the minimum $F$ for which a 
moving solution exists, and 
$F_{c0}$, as should be expected.  Thus, if $F$ is increased 
above $F_{c}^{\uparrow}$ and back down there will be hysteresis.  If
$F$ is decreased below $F_{c}^{\downarrow}$ so that the system stops and is
then increased back up again, the behavior can be complicated as
$\rho(f)$ will generally {\it not} be continuous.  Depending on the parameters
(including $D(\Delta)$) and the history, the depinning transition on the
second increase of $F$ can either be of the hybrid type discussed above
with an exponentially growing characteristic size, $\hat{s}$, of large avalanches which
diverges at a {\it new} $F^{\uparrow}_{c}$; or a strongly discontinuous
transition with $\Gamma$ jumping from below $\Gamma_{c}$ to above
discontinuously and no extensive number of precusor
large avalanches.

We have found that in a simple infinite-range model of depinning with 
dynamic stress overshoots, rather subtle behavior can occur.  
For large overshoots, a hybrid transition exists when the force is increased adiabatically. This is characterized by
exponentially-large, almost-deterministic avalanches 
followed by runaway at a critical force to a non-zero velocity
state; this moving state persists to a {\it lower} critical force as the force is decreased
back down.  Various history dependent effects can occur in this large overshoot regime.  
For
smaller stress overshoots, there is a reversible critical 
transition similar to that in the absence of overshoots.  These
two regimes are separated by a ``tricritical'' point which represents  a 
new universality class.  

 All of this 
behavior should obtain for a broad class of infinite-range models,
including ones with locally underdamped relaxation caused by inertia [12,13].  
But the
crucial question is: what of this behavior persists in finite-dimensional 
models?  For large stress overshoots the depinning
is surely of a different character than in purely dissipative systems,
although what aspects of it might resemble the infinite-range model is far 
from clear.  For small stress overshoots, it has been argued elsewhere
that the critical force on increasing $M$ is generically 
{\it less} than $F_{c0}$ in finite-dimensional models with 
long-range interactions [14]. This is  the relevant situation for fronts of planar cracks propagating through heterogeneous brittle
materials and  one would guess that the same should hold more generally.
But whether this implies that the transition immediately changes character when {\it any} stress overshoots are present --
and thus presumably that there is no tricritical point -- or whether 
dissipative-like critical behavior can still exist with small
overshoots, we must leave as an open question. 

The authors would like to thank Ron
Maimon and Cristina Marchetti for useful discussions.  This work has been supported in part by the National Science Foundation via DMR-9630064, DMR-9976621, and DMR-9809363.

\end{document}